

\font\titlefont = cmr10 scaled\magstep 4
\font\sectionfont = cmr10
\font\littlefont = cmr5
\font\eightrm = cmr8

\def\ss{\scriptstyle}
\def\sss{\scriptscriptstyle}

\newcount\tcflag
\tcflag = 0  
\ifnum\tcflag = 0 \magnification = 1200 \fi  

\global\baselineskip = 1.2\baselineskip
\global\parskip = 4pt plus 0.3pt
\global\abovedisplayskip = 18pt plus3pt minus9pt
\global\belowdisplayskip = 18pt plus3pt minus9pt
\global\abovedisplayshortskip = 6pt plus3pt
\global\belowdisplayshortskip = 6pt plus3pt


\def\endignore{}
\def\ignore #1\endignore{}

\newcount\dflag
\dflag = 0


\def\monthname{\ifcase\month
\or January \or February \or March \or April \or May \or June%
\or July \or August \or September \or October \or November %
\or December
\fi}

\newcount\dummy
\newcount\minute  
\newcount\hour
\newcount\localtime
\newcount\localday
\localtime = \time
\localday = \day

\def\advanceclock#1#2{ 
\dummy = #1
\multiply\dummy by 60
\advance\dummy by #2
\advance\localtime by \dummy
\ifnum\localtime > 1440 
\advance\localtime by -1440
\advance\localday by 1
\fi}

\def\settime{{\dummy = \localtime%
\divide\dummy by 60%
\hour = \dummy
\minute = \localtime%
\multiply\dummy by 60%
\advance\minute by -\dummy
\ifnum\minute < 10
\xdef\spacer{0} 
\else \xdef\spacer{}
\fi %
\ifnum\hour < 12
\xdef\ampm{a.m.} 
\else
\xdef\ampm{p.m.} 
\advance\hour by -12 %
\fi %
\ifnum\hour = 0 \hour = 12 \fi
\xdef\timestring{\number\hour : \spacer \number\minute%
\thinspace \ampm}}}



\def\endtitle{}
\def\title#1\endtitle{\vskip.5in\titlefont
\global\baselineskip = 2\baselineskip
#1\vskip.4in
\baselineskip = 0.5\baselineskip\rm}

\def\endauthors{}
\def\authors#1\endauthors{#1}

\def\endabstract{}
\def\abstract#1\endabstract{\vskip .3in%
\centerline{\sectionfont\bf Abstract}%
\vskip .1in
\noindent#1}

\newcount\nsection
\newcount\nsubsection

\def\section#1{\global\advance\nsection by 1
\nsubsection=0
\bigskip\noindent\centerline{\sectionfont \bf \number\nsection.\ #1}
\bigskip\rm\nobreak}

\def\subsection#1{\global\advance\nsubsection by 1
\bigskip\noindent\sectionfont \sl \number\nsection.\number\nsubsection)\
#1\bigskip\rm\nobreak}


\def\appendix#1#2{\bigskip\noindent%
\centerline{\sectionfont \bf Appendix #1.\ #2}
\bigskip\rm\nobreak}


\newcount\nref
\global\nref = 1

\def\ref#1#2{\xdef #1{[\number\nref]}
\ifnum\nref = 1\global\xdef\therefs{\noindent[\number\nref] #2\ }
\else
\global\xdef\oldrefs{\therefs}
\global\xdef\therefs{\oldrefs\vskip.1in\noindent[\number\nref] #2\ }%
\fi%
\global\advance\nref by 1
}

\def\listrefs{\bigskip\section{References}\therefs}


\newcount\nfoot
\global\nfoot = 1

\def\foot#1#2{\xdef #1{(\number\nfoot)}
\footnote{${}^{\number\nfoot}$}{\eightrm #2}
\global\advance\nfoot by 1
}


\newcount\nfig
\global\nfig = 1

\def\fig#1{\xdef #1{(\number\nfig)}
\global\advance\nfig by 1
}


\newcount\cflag
\newcount\nequation
\global\nequation = 1
\def\eqlabel{(1)}

\def\nexteqno{\ifnum\cflag = 0
\global\advance\nequation by 1
\fi
\global\cflag = 0
\xdef\eqlabel{(\number\nequation)}}

\def\lasteqno{\global\advance\nequation by -1
\xdef\eqlabel{(\number\nequation)}}

\def\label#1{\xdef #1{(\number\nequation)}
\ifnum\dflag = 1
{\escapechar = -1
\xdef\draftname{\littlefont\string#1}}
\fi}

\def\clabel#1#2{\xdef\eqlabel{(\number\nequation #2)}
\global\cflag = 1
\xdef #1{\eqlabel}
\ifnum\dflag = 1
{\escapechar = -1
\xdef\draftname{\string#1}}
\fi}

\def\cclabel#1#2{\xdef\eqlabel{#2)}
\global\cflag = 1
\xdef #1{\eqlabel}
\ifnum\dflag = 1
{\escapechar = -1
\xdef\draftname{\string#1}}
\fi}


\def\eeq{}

\def\eqnn #1\eeq{$$ #1 $$}

\def\eq #1\eeq{
\ifnum\dflag = 0
{\xdef\draftname{\ }}
\fi 
$$ #1
\eqno{\eqlabel \rlap{\ \draftname}} $$
\nexteqno}




\def\eeol{& \eqlabel \rlap{\ \draftname}
\nexteqno
\xdef\draftname{\ }}

\def\eolnn{\cr
\global\cflag = 0
\xdef\draftname{\ }}


\def\eqa #1\eeq{
\ifnum\dflag = 0
{\xdef\draftname{\ }}
\fi 
$$ \eqalignno{ #1 } $$
\global\cflag = 0}


\def\ie{{\it i.e.\/}}


\def\plb#1#2#3{{\it Phys.\ Lett.} {\bf #1B} (19#2) #3}

\def\prd#1#2#3{{\it Phys.\ Rev.} {\bf D#1} (19#2) #3}

\def\prl#1#2#3{{\it Phys.\ Rev.\ Lett.} {\bf #1} (19#2) #3}


\global\nulldelimiterspace = 0pt



\def\frac#1#2{{{#1} \over {#2}}\,}  



\def\Dsl{\hbox{/\kern-.6700em\it D}} 
\def\dsl{\hbox{/\kern-.5300em$\partial$}}
\def\pxpsl{\hbox{/\kern-.5600em$p$}}
\def\ssl{\hbox{/\kern-.5300em$s$}}
\def\epssl{\hbox{/\kern-.5100em$\epsilon$}}
\def\delsl{\hbox{/\kern-.6300em$\nabla$}}
\def\lxpsl{\hbox{/\kern-.4300em$l$}}
\def\elxpsl{\hbox{/\kern-.4500em$\ell$}}
\def\kxpsl{\hbox{/\kern-.5100em$k$}}
\def\qxpsl{\hbox{/\kern-.5000em$q$}}
\def\sla#1{\raise.15ex\hbox{$/$}\kern-.57em #1}



\def\roughly#1{\mathrel{\raise.3ex\hbox{$#1$\kern-.75em\lower1ex\hbox{$\sim$}}}}
\def\lsim{\roughly<}






\def\ssl{{\sss L}}







\overfullrule=0pt


\def\ab{{\alpha\beta}}
\def\oq{{\omega_q}}
\def\ks{K_{\sss S}}
\def\ss{\scriptstyle}
\def\bentarrow{{\raise1.1ex\hbox{\rlap{$\vert$}}\kern-.2em\rightarrow}}
\def\barp{{\raise.35ex\hbox{${\sss (}$}}---{\raise.35ex\hbox{${\sss )}$}}}
\def\bdbarp{\hbox{$B_d$\kern-1.4em\raise1.4ex\hbox{\barp}}}
\def\bsbarp{\hbox{$B_s$\kern-1.4em\raise1.4ex\hbox{\barp}}}
\def\oabij{\omega_\ab^{ij}}


\rightline{DAPNIA/SPP 94-06}
\rightline{NSF-PT-94-2}
\rightline{UdeM-LPN-TH-94-189}
\rightline{March 1994}
\vskip .2in

\title
\centerline{Determining the Quark Mixing Matrix}
\centerline{From CP-Violating Asymmetries}
\endtitle

\authors
\centerline{R. Aleksan,${}^a$, B. Kayser${}^b$ and D. London${}^c$}
\vskip .15in
\centerline{\it ${}^a$ DAPNIA, CE Saclay}
\centerline{\it F-91191 Gif-sur-Yvette Cedex, France}
\vskip .1in
\centerline{\it ${}^b$ Division of Physics, National Science Foundation}
\centerline{\it 4201 Wilson Blvd., Arlington, VA 22230 USA}
\vskip .1in
\centerline{\it ${}^c$ Laboratoire de Physique Nucl\'eaire, Universit\'e de
Montr\'eal}
\centerline{\it C.P. 6128, Montr\'eal, Qu\'ebec, CANADA, H3C 3J7.}
\endauthors

\abstract
If the Standard Model explanation of CP violation is correct, then
measurements of CP-violating asymmetries in $B$ meson decays can in
principle determine the entire quark mixing matrix.
\endabstract
\vskip1truecm


\ref\neutralBCP{For reviews, see, for example, Y. Nir and H.R. Quinn in
``$B$ Decays,'' ed.\ S. Stone (World Scientific, Singapore, 1992), p.\ 362;
I. Dunietz, {\it ibid} p.\ 393.}
According to the Standard Model (SM), CP violation arises from the fact
that in the Cabibbo-Kobayashi-Maskawa (CKM) quark mixing matrix, some of
the elements are not real. This picture of CP violation will be incisively
tested in neutral $B$ meson decays, where some of the CP-violating
asymmetries can yield theoretically clean information on the phases of
various products of CKM elements \neutralBCP.

While the SM does not predict the elements of the CKM matrix, $V$, in
detail, it does require that $V$ be unitary. This requirement is, as we
shall see, a very powerful constraint. It implies, among other things, that
any pair of columns, or any pair of rows, of $V$ be orthogonal. Thus,
assuming that there are three generations of quarks, so that $V$ is a
$3\times 3$ matrix, we have the six orthogonality conditions
\label\orthogonality
\eq
\eqalign{
\sum_{\alpha=1}^3 V_{\alpha i}V_{\alpha j}^* & = 0~,~~i\ne j, \cr
\sum_{i=1}^3 V_{\alpha i}V_{\beta i}^* & = 0~,~~\alpha\ne \beta. \cr}
\eeq
Here and hereafter, Greek subscripts run over the up-type quarks $u,c$ and
$t$, while Latin ones run over the down-type quarks $d,s$ and $b$. It is
often useful to picture each of Eqs.~\orthogonality\ as the statement that
the ``unitarity triangle'' in the complex plane whose sides are the terms
in the equation is closed. The six unitarity triangles corresponding to
Eqs.~\orthogonality\ are depicted, somewhat schematically, in Fig.~1. We
refer to each of these triangles by naming the columns or rows whose
orthogonality it represents. As shown in Fig.~1, in the $ds$, $sb$, $uc$
and $ct$ triangles, one leg is known empirically to be short compared to
the other two, so that the angle opposite the short leg is small.

\ref\angles{See, for example, C. Jarlskog in `` CP Violation,'' ed.\ C.
Jarlskog (World Scientific, Singapore, 1989), p.\ 3; J. Rosner in ``$B$
Decays,'' ed.\ S. Stone (World Scientific, Singapore, 1992), p.\ 312.}
\ref\KLAlonger{A more complete account of our results will be presented
elsewhere.}
Apart from an extra $\pi$ and a possible minus sign, each of the angles in
the unitarity triangles is just the relative phase of the two adjacent
legs. Let
\eq
\label\omegadef
\omega_\ab^{ij} \equiv arg(V_{\alpha i} V_{\alpha j}^* / V_{\beta i}
V_{\beta j}^*),
\eeq
with $\alpha\ne\beta$ and $i\ne j$, be the relative phase of the leg
involving the up-type quark $\alpha$ (the ``$\alpha$ leg'') and the leg
involving the up-type quark $\beta$ (the ``$\beta$ leg'') in the $ij$
column triangle. Since
$arg(V_{\alpha i} V_{\alpha j}^* / V_{\beta i} V_{\beta j}^*) =
arg(V_{\alpha i} V_{\beta i}^* / V_{\alpha j} V_{\beta j}^*)$,
$\omega_\ab^{ij}$ is also the relative phase of the $i$ and $j$ legs in the
$\alpha\beta$ row triangle. At most, four of the $\omega_\ab^{ij}$ can be
independent, since four parameters, usually taken to be three mixing angles
and one complex phase, are sufficient to fully determine $V$ \angles.
Indeed, it is easy to show \KLAlonger\ that, mod $2\pi$, any
$\omega_\ab^{ij}$ is a simple linear combination of, for example, the four
phases $\omega_{tu}^{bd}$, $\omega_{ct}^{bd}$, $\omega_{tu}^{sb}$ and
$\omega_{ct}^{sb}$. Let us now demonstrate that the phase of any
phase-convention-independent product of CKM elements (that is, the quantity
probed by the CP asymmetry in any neutral $B$ decay) is a linear
combination of these four phases, with integer coefficients. To show this,
it is convenient to work in the phase convention where all elements of $V$
are real and positive, except for $V_{ud}$, $V_{us}$, $V_{cd}$ and
$V_{cs}$. In this phase convention,
\label\phasecon
\eq
\eqalign{
\omega_{tu}^{bd} & \equiv arg (V_{tb} V_{td}^* / V_{ub} V_{ud}^* ) = arg
(V_{ud}) \cr
\omega_{tu}^{sb} & \equiv arg (V_{ts} V_{tb}^* / V_{us} V_{ub}^* ) = -arg
(V_{us}) \cr
\omega_{ct}^{bd} & \equiv arg (V_{cb} V_{cd}^* / V_{tb} V_{td}^* ) = -arg
(V_{cd}) \cr
\omega_{ct}^{sb} & \equiv arg (V_{cs} V_{cb}^* / V_{ts} V_{tb}^* ) = arg
(V_{cs}). \cr}
\eeq
Now, suppose $P$ is some phase-convention-independent product of CKM
elements. In our present phase convention, the phase of $P$, $\omega$, is
given by
\label\phasesum
\eq
\omega = n_{ud} \, arg(V_{ud}) + n_{us} \, arg(V_{us}) + n_{cd} \,
arg(V_{cd}) + n_{cs} \, arg(V_{cs}).
\eeq
Here, $n_{\alpha i}$ is the number of factors of $V_{\alpha i}$ appearing
in $P$, minus the number of factors of $V_{\alpha i}^*$. From
Eqs.~\phasesum\ and \phasecon, $\omega$ is given in the present phase
convention by
\label\omegasum
\eq
\omega = n_{ud} \, \omega_{tu}^{bd} - n_{us} \, \omega_{tu}^{sb}
       - n_{cd} \, \omega_{ct}^{bd} + n_{cs} \, \omega_{ct}^{sb}~.
\eeq
That is,
\label\omegalincomb
\eq
\omega = \sum_{q=1}^4 n_q \, \oq~,
\eeq
where the $\oq$ are the phases $\omega_{tu}^{bd}$, etc., appearing in
Eq.~\omegasum, and the $n_q$ are integers which are known for any given
CKM product $P$. Since $\omega$ is phase-convention-independent, and, as one
may easily verify, so are the $\omega_\ab^{ij}$, the relation
\omegalincomb\ must hold in {\it any} phase convention.

We see that the four phases $\oq$ form a complete set of variables for the
description of CP violation in $B$ decay. Moreover, through
Eq.~\omegalincomb\ the phases $\omega$ that are probed in the $B$
experiments are related {\it very simply} to the $\oq$. In contrast, when
$V$ is treated exactly, the phases $\omega$ are quite complicated functions
of the quark mixing angles and complex phase factor often used to
parametrize $V$. Thus, it appears useful to think of measurements of CP
violation in $B$ decays as probes of the variables $\oq$.

Imagine that through observation of CP-violating asymmetries in $B$ decays
we have determined the four phases $\oq$. Let us show that from these
phases we can reconstruct the entire CKM matrix!

To determine the {\it phases} of the CKM elements from the $\oq$, we must
first choose a phase convention, because the phases of individual
$V_{\alpha i}$ change with quark-field phase redefinitions, while the $\oq$
do not. Let us adopt the phase convention defined just before
Eqs.~\phasecon. In this convention, the only $V_{\alpha i}$ which are not
real and positive are the four whose phases are given in terms of the $\oq$
by Eqs.~\phasecon. Thus, all nontrivial phases in $V$ are determined by the
$\oq$.

To see how the {\it magnitudes} of the CKM elements are determined by the
$\oq$, let us first note that, according to the law of sines, the ratio
between the $\alpha$ and $\beta$ legs of the $ij$ triangle is given by
\eq
{ \vert V_{\alpha i} V_{\alpha j}^* \vert \over
  \vert V_{\beta i} V_{\beta j}^* \vert} =
{ \left\vert \sin \omega_{\beta\gamma}^{ij} \right\vert \over
\left\vert \sin \omega_{\gamma\alpha}^{ij} \right\vert }~.
\eeq
Here and in the following relation, $\alpha\beta\gamma$ is some cyclic
permutation of $uct$, and $ijk$ of $dsb$. Applying the law of sines to the
$jk$ and $ki$ triangles as well, we find that
\label\vckmratio
\eqa
{ \vert V_{\alpha i} \vert^2 \over \vert V_{\beta i} \vert^2 } & =
\left\vert { V_{\alpha i} V_{\alpha j}^* \over V_{\beta i} V_{\beta j}^* }
\right\vert
\left\vert { V_{\beta j} V_{\beta k}^* \over V_{\alpha j} V_{\alpha k}^* }
\right\vert
\left\vert { V_{\alpha k} V_{\alpha i}^* \over V_{\beta k} V_{\beta i}^* }
\right\vert \eolnn
& = \left\vert {\sin \omega_{\beta\gamma}^{ij} \over
\sin \omega_{\gamma\alpha}^{ij} } \right\vert
\;\; \left\vert {\sin \omega_{\gamma\alpha}^{jk} \over
\sin \omega_{\beta\gamma}^{jk} } \right\vert
\;\; \left\vert {\sin \omega_{\beta\gamma}^{ki} \over
\sin \omega_{\gamma\alpha}^{ki} } \right\vert ~. \eeol
\eeq
The phase angles in the column triangles appearing on the right-hand side
of this relation are all known once the $\oq$ are known. Thus, given the
$\oq$, Eq.~\vckmratio\ with $\alpha=c$ and $\beta=u$ determines $\vert
V_{ci} \vert^2 / \vert V_{ui} \vert^2 \equiv a_i$, and with $\alpha=t$ and
$\beta=u$ determines  $\vert V_{ti} \vert^2 / \vert V_{ui} \vert^2 \equiv
b_i$. If we then impose the unitarity constraint on the $i^{th}$ column of
$V$,
\label\unitarity
\eq
\vert V_{ui} \vert^2 + \vert V_{ci} \vert^2 + \vert V_{ti} \vert^2 = 1~,
\eeq
we obtain $\vert V_{ui}\vert^2$:
\eq
\vert V_{ui} \vert^2 = {1 \over 1 + a_i + b_i } ~.
\eeq
{}From $a_i$ and $b_i$, the remaining magnitudes in the $i^{th}$ column,
$\vert V_{ci}\vert^2$ and $\vert V_{ti}\vert^2$, then follow immediately. In
this way, the phases $\oq$ determine the magnitudes of all the elements of
$V$.

It is worth emphasizing the crucial role played here by unitarity. Naively,
one might guess that, by determining all the interior angles in the
unitarity triangles, the phases $\oq$ would fix the {\it shapes} of these
triangles, but not their {\it sizes}. There would then be nothing to set
the scale for the magnitudes of individual CKM elements. However, in
reality this scale is set by unitarity via Eq.~\unitarity.

\ref\jarlskog{Just as four independent phases $\oq$ fully determine $V$, so
do four independent magnitudes $\vert V_{\alpha i}\vert$ plus one sign
which fixes the orientation of the unitarity triangles. See C. Jarlskog,
Ref.~\angles.}
Since the four $\oq$ fully determine $V$, and we know four {\it independent}
parameters are required to do that, it is now obvious that the $\oq$ are
independent parameters. Attempts to find relations among them within the SM
would prove fruitless. Of course, by using the linear relations that permit
us to express some of the nine $\oabij$ in terms of others \KLAlonger, we
can replace the four independent $\oq$ of Eq.~\omegasum\ by any other set
of four independent phases in the unitarity triangles \jarlskog.

\ref\wolfenstein{L. Wolfenstein, \prl{51}{83}{1945}.}
Is it feasible for $B$ decay experiments to determine the four independent
phases $\oq$, and through them the entire CKM
matrix? To explore this question, it is illuminating to choose the four
independent $\oq$, not as the phases which appear in Eq.~\omegasum, but as
\label\omegaset
\eq
\omega_1 \equiv \omega_{tu}^{bd}~,~~~~
\omega_2 \equiv \omega_{ct}^{bd}~,~~~~
\omega_3 \equiv \omega_{ct}^{sb}~,~~~~
\omega_4 \equiv \omega_{uc}^{ds}~.
\eeq
Referring to Fig.~1, and recalling the rough magnitudes of the CKM elements
summarized by the Wolfenstein approximation \wolfenstein, we see that
$\omega_1$ and $\omega_2$ may both be large (\ie, far from 0 and $\pm\pi$),
but  $\omega_3=\pm(\pi-\epsilon)$ with $\epsilon \lsim 0.05$ radians, and
$\omega_4=\pm(\pi-\epsilon')$ with $\epsilon' \lsim 0.003$ radians. In
Table 1, we list for purposes of illustration a set of $B$ decay modes
whose study could in principle determine all four $\oq$ of Eq.~\omegaset,
with no ambiguities.

\ref\growyler{ The decays $B^\pm \to D K^\pm$, followed by $D\to K^+ K^-$,
cleanly yield CKM phase information even though they involve charged parent
$B$ mesons. See M. Gronau and D. Wyler, \plb{265}{91}{172}.}
\ref\psiphi{ The decays $\bsbarp\to\Psi\phi$ are expected to have a CP
asymmetry of only a few percent, but a relatively large branching ratio of
$\sim 10^{-3}$, facilitating their study. See I. Dunietz, Fermilab preprint
Fermilab-Conf-93/90-T, to appear in the Proceedings of the Workshop on $B$
Physics at Hadron Accelerators, Snowmass, Colorado, June 21-July 2, 1993.}
\topinsert
$$\vbox{\tabskip=0pt \offinterlineskip
\halign to \hsize{\strut#& #\tabskip 1em plus 2em minus .5em&
\hfil$#$\hfil &#&\hfil$#$\hfil &#\tabskip=0pt\cr
\noalign{\hrule}\noalign{\smallskip}\noalign{\hrule}\noalign{\medskip}
&& \hbox{Decay Mode} && \hbox{Quantity Determined by}&\cr
&& \omit && \hbox{Observed CP Asymmetry} &\cr
\noalign{\medskip}\noalign{\hrule}\noalign{\medskip}
&& \bdbarp\ \to \pi^+\pi^- && \sin 2\omega_{tu}^{bd} &\cr
\noalign{\smallskip}
&& \bdbarp\ \to \Psi\ks && \sin 2\omega_{ct}^{bd} &\cr
\noalign{\smallskip}
&& B^\pm \to D K^\pm && \sin^2 (\omega_{uc}^{bd} + \omega_{uc}^{ds} ) &\cr
&& ~\quad\qquad\qquad\bentarrow K^+ K^- && \omit &\cr
\noalign{\smallskip}
&& \bsbarp\ \to D_s^\pm K^\mp &&
\sin^2 (\omega_{uc}^{bd} - 2\omega_{ct}^{sb} + \omega_{uc}^{ds} ) \} &\cr
\noalign{\smallskip}
&& \bsbarp\ \to \Psi\phi && \sin 2 \omega_{ct}^{sb} &\cr
\noalign{\medskip}\noalign{\hrule}\noalign{\smallskip}\noalign{\hrule}
}}$$
\medskip
\noindent {\eightrm Table 1. An illustrative complete set of $\ss B$ decay
experiments \growyler, \psiphi.}
\endinsert

\ref\paris{We thank Paris Sphicas for pointing out that this constraint can
help resolve ambiguities.}
\ref\snyquinn{The decays $\bsbarp\to D_s^\pm K^\mp$ need be used only to
resolve a discrete ambiguity. This same discrete ambiguity can also be
resolved by determining $\cos 2\omega_{tu}^{bd}$ through study of the
decays $\bdbarp\to\rho\pi$. How these decays determine this quantity is
explained in A.E. Snyder and H.R. Quinn, \prd{48}{93}{2139}.}
\ref\ambiguities{Ambiguities remain for special, isolated values of the
phases, but such values are unlikely.}
As Table 1 illustrates, the quantities determined by CP asymmetries in $B$
decays are not precisely the phases of products of CKM elements, but
trigonometric functions of these phases. These functions leave the phases
themselves discretely ambiguous. However, after a lengthy but
straightforward analysis \KLAlonger, it is found that, together, the
measurements in Table 1, supplemented by the constraint that the angles in
any triangle add up to $\pi$ \paris, determine the four phases \omegaset\
without ambiguities \snyquinn, \ambiguities.

\ref\rosner{See, for example, J. Rosner, Ref.~\angles.}
\ref\vcdinput{If one is willing to use $\vert V_{cd} \vert$ as an input,
then one can also find $\vert V_{ub}/V_{cb} \vert$ using the relation
$\vert V_{ub}/V_{cb} \vert / \vert V_{cd}/V_{ud} \vert = \sin\beta /
\sin\alpha$, obtained by applying the law of sines to the $bd$ triangle. If
one is further willing to assume $\vert V_{tb} \vert \simeq 1$, then one
can find $\vert V_{td}/V_{cb} \vert$ using the relation $\vert V_{td}
V_{tb} \vert /\vert V_{cd} V_{cb} \vert = \sin\gamma / \sin\alpha$,
obtained similarly.}
Clearly, the determination of the larger phases $\omega_1$, $\omega_2$ and
$\omega_3$ may be feasible, but, given that $\epsilon'\lsim 0.003$, the
determination of $\omega_4$ would be extremely difficult, if not
impossible. For example, to determine $\omega_4\equiv\omega_{uc}^{ds}$ from
the decays in Table 1, we would first have to use the first two of these
decays to very accurately fix $\omega_{tu}^{bd}$ and $\omega_{ct}^{bd}$,
from which $\omega_{uc}^{bd}$ follows. We would then have to use the decays
$B^\pm \to DK^\pm \to (K^+K^-)K^\pm$ to determine $\sin^2(\omega_{uc}^{bd}
+ \omega_{uc}^{ds})$ to better than $\pm 0.001$! Suppose, then, that
$\omega_4$ proves to be beyond reach. Can we learn interesting things about
the CKM matrix, and in particular about the magnitudes of its elements,
from a knowledge of $\omega_1$, $\omega_2$ and $\omega_3$ alone? Indeed we
can. Neglecting $\epsilon$ and $\epsilon'$ compared to $\omega_1$ and
$\omega_2$, we find from Eq.~\vckmratio\ that
\label\vubvcb
\eq
\left\vert{V_{ub} \over V_{cb}} \right\vert^2 \simeq
{\sin \beta \sin\epsilon \over \sin\alpha \sin\gamma }~.
\eeq
Here, $\alpha\equiv\pi-\vert \omega_1\vert$, $\beta\equiv\pi-\vert
\omega_2\vert$, and $\gamma=\pi-\alpha-\beta$ are the same (positive)
interior angles of the $bd$ triangle commonly denoted by these symbols in
the literature \neutralBCP. Similarly, from the analogue of Eq.~\vckmratio\
for the ratio of CKM elements in one row, we find that
\label\vtdvts
\eq
\left\vert{V_{td} \over V_{ts}} \right\vert^2 \simeq
{\sin \gamma \sin\epsilon \over \sin \alpha \sin\beta }~.
\eeq
Since $\vert V_{cb}\vert$ is known and $\vert V_{ts}\vert \simeq \vert
V_{cb} \vert$  \rosner, Eqs.~\vubvcb\ and \vtdvts\ determine $\vert V_{ub}
\vert$ and $\vert V_{td} \vert$ in terms of CP-violating angles. This is
interesting, because these very small CKM elements are difficult to
determine in other ways. Note that their determination via Eqs.~\vubvcb\
and \vtdvts\ has the advantage of being free of theoretical hadronic
uncertainties \vcdinput, apart from those needed to fix $\vert
V_{cb}\vert$.

{}From the analogue of Eq.~\vckmratio\ for CKM elements in one row, we find
the relation
\eq
\left\vert {V_{us} \over V_{ud}} \right\vert^2 \simeq
{\sin \alpha \sin\epsilon \over \sin\beta \sin\gamma}~,
\eeq
expressing the square of the Cabibbo angle in terms of CP-violating angles.
The Cabibbo angle is, of course, very well known, so this relation can
serve as a good test of the SM explanation of CP violation.

Conceptually, it is very interesting that the entire CKM matrix can in
principle be determined by CP-violating $B$-decay asymmetries alone. This
implies that these asymmetries can serve as an incisive probe of the
structure of the matrix responsible, according to the SM, for CP violation.
Perhaps these asymmetries can even be a practical source of significant
information on $\vert V_{ub}/V_{cb} \vert$ and $\vert V_{td}/V_{ts} \vert$.

\bigskip
\centerline{\bf Acknowledgments}
\bigskip
It is a pleasure to thank A. Ali, I. Dunietz, J. Rosner and P. Sphicas for
helpful conversations. Two of us (BK and DL) are grateful for the
hospitality of Fermilab, where part of this work was done. One of us (BK)
is also grateful for the hospitality of DESY, where another part was done.
This research was partially funded by the N.S.E.R.C.\ of Canada and les
Fonds F.C.A.R.\ du Qu\'ebec.

\listrefs

\vfill
\noindent
Figure Caption:

\noindent
The unitarity triangles. To the left of each triangle is indicated the pair
of columns, or of rows, whose orthogonality this closed triangle expresses.

\bye